\documentclass{article}
\usepackage[latin9]{inputenc}
\usepackage{amsmath}
\usepackage{amssymb}

\makeatletter

\newcommand{\noun}[1]{\textsc{#1}}

\makeatother

\begin{document}
%%% remove comment delimiter ('%') and select language if required%\selectlanguage{spanish} 

\noindent \noun{T}\emph{\noun{he Stephani Universe, k-essence and
strings in the 5-th dimension}}\medskip{}

\noindent G. B. Tupper%
\footnote{ $ $ $Email$ $ $ $gary.tupper@uct.ac.za$ %
}$^{2}$, M Marais%
\footnote{ $ $ $Email$ $ $ $markmarais31@gmail.com$ %
} and J A Helayël$^ {}$%
\footnote{ $ $ $Email$ $ $ $helayel@cbpf.br$ %
}

\noindent \textit{$^{1}$Department of Physics, University of Cape
Town, Private Bag, Rondebosch 7700, South Africa }

\noindent \textit{Associate member, NITheP}

\noindent \textit{$^{2}$Cape Peninsula University of Technology,
Symphony Way, Bellville 7530, South Africa }

\noindent \textit{$^{3}$Centre for Brazilian Research in Physics,}
\textit{Rua Dr. Xavier Sigaud, 150 Urca, Rio de Janeiro, RJ, Brasil\medskip{}
 }

\noindent ABSTRACT: The Stephani universe is an inhomogeneous alternative
to $\Lambda{\rm CDM}$ .We show that the exotic fluid driving the
Stephani exact solution of Einstein's equations is an unusual form
of k-essence that is linear in ``velocity''. Much as the Stephani
universe can be embedded into (a section of) flat 5-d Minkowski space-time,
we show that the k-essence obtains through dimensional reduction of
a 5-d strongly coupled non-linear ``electrodynamics'' that, in the
empty Stephani universe, is similar to the Nielsen-Olesen field theory
of the dual string.\medskip{}

\paragraph{ Introduction}

\noindent The discovery {[}1{]} that distant supernova are dimmer
than they would be in a Einstein -- de Sitter universe has forced
a fresh contemplation of issues almost as old as general relativity
itself. If accelerated Hubble expansion driven by a cosmological constant$\Lambda$
is the explanation, one is left the difficult task of explaining why$\Lambda$
is infinitesimally small in natural units if the Planck mass, and
just such as to reveal itself in the present epoch {[}2{]}.

\noindent A much discussed alternative is to assume that the cosmological
constant vanishes and the acceleration is due to the nonlinearity
of the Einstein equations: since the present universe is only homogeneous
and isotropic on average, averaging of the Einstein equations will
yield the usual FRW model equations plus corrections from `back-reaction'
{[}3{]}. It is easy to realise this by constructing spherical LTB
metrics that can account for the supernova data {[}4{]}, however such
models assume a co-moving co-ordinate system and rely upon significant
shear. Recalling that the inhomogeneity in the solar system are far
larger than the cosmological one yet are readily treated by post-Newtonian
approximation suggests that corrections to the usual linearly perturbed
FRW model will be similarly small, and indeed this proves to be the
case {[}5{]}.

\noindent Still, the large-scale homogeneity of the universe is much
less observationally secure than its isotropy {[}6{]}, so suggesting
the simplest inhomogeneous but isotropic and shear-free generalization
of the FRW metric {[}6{]}: 
\begin{equation}
ds^{2}=N\left(t,r\right)^{2}dt^{2}-R\left(t,r\right)^{2}d\vec{x}^{2}\label{1.1)}
\end{equation}
Assuming a perfect fluid source $T_{\mu\nu}=\left(\rho+p\right)u_{\mu}u_{\nu}-g_{\mu\nu}p$,
the resulting Einstein equations%
\footnote{ $We\ use\ units$ $8\pi G=c=1$ %
} we first solved by Wyman {[}7{]}; the vanishing of the off-diagonal
components of the source in co-moving co-ordinates provides 
\begin{equation}
\begin{array}{l}
{\frac{R_{,t}}{NR}=\frac{\dot{R}}{R}=H\left(t\right)}\\
\\
{N=\frac{R_{,t}}{H\left(t\right)R}}
\end{array}\label{1.2)}
\end{equation}
Herein e.g. $\dot{R}=N^{-1}R_{,t}$ indicates the proper time derivative
and$\ H\left(t\right)$ is a function of integration. Further, as\ $T_{i}^{j}=-p\delta_{i}^{j}$
implies\ $G_{r}^{r}=G_{\theta}^{\theta}=G_{\phi}^{\phi}$ , one is
lead to the ``pressure isotropy equation'': 
\begin{equation}
R''-2\frac{R'^{2}}{R}-\frac{R'}{r}=\frac{1}{2}f\left(r\right)\label{1.3)}
\end{equation}
The ``primes'' here denote partial derivatives with respect to\ $r$
and\ $f\left(r\right)$ is another integration function. The remaining
Einstein equations can be expressed in the Friedman-like form 
\begin{equation}
\begin{array}{l}
{3H^{2}=\rho+\frac{1}{R^{2}}\left[\frac{2R''}{R}-\left(\frac{R'}{R}\right)^{2}+\frac{4}{r}\frac{R'}{R}\right]}\\
{\dot{\rho}+3\frac{\dot{R}}{R}\left[\rho+p\right]=0}
\end{array}\label{1.4)}
\end{equation}
Wyman's objective was to obtain solutions for a baratropic equation
of state\ $p=p\left(\rho\right)$ so that he excluded a solution
that would later be rediscovered by Stephani {[}8{]} 
\begin{equation}
\, f=0\,,\quad H=\frac{a_{,t}}{a},\quad R=\frac{a\left(t\right)}{1+{ka\left(t\right)r^{2}\mathord{\left/{\vphantom{ka\left(t\right)r^{2}4}}\right.\kern -\nulldelimiterspace}4}}=aN\label{1.5)}
\end{equation}
The corresponding energy density and pressure follow as 
\begin{equation}
\rho=3\left[H^{2}+{k\mathord{\left/{\vphantom{ka}}\right.\kern -\nulldelimiterspace}a}\right],\quad p=-\rho-\frac{1}{3}\frac{1+{ka\left(t\right)r^{2}\mathord{\left/{\vphantom{ka\left(t\right)r^{2}4}}\right.\kern -\nulldelimiterspace}4}}{H}\rho_{,t}\label{ZEqnNum850376}
\end{equation}
That is to say the energy density is homogeneous while the pressure
is inhomogeneous. Thus while the Stephani model has been considered
as an alternative to the$\Lambda CDM$model {[}9{]}, its viability
is obscured by the question: what is the nature of the perfect fluid
source having these unusual properties.

\noindent In this paper we will provide an answer to the aforementioned
question: the source is a particular case of ``k-essence'' {[}10{]}
having a Lagrangian density that is linear in the `velocity'. Moreover,
just as the Stephani metric is exceptional in that it can be embedded
into 5-dimensional Minkowski space {[}8{]}, so too can the k-essence
source be lifted to a 5-dimensional nonlinear `electrodynamics'.

\noindent The remainder of this paper is organised as follows: in
Section 2 we briefly review and reformulate k-essence in a way that
makes the choice of Lagrangian density yielding \eqref{ZEqnNum850376}
self-evident. Then in Section 3 we show how general k-essence models
can be obtained by dimensional reduction from 5-dimensional nonlinear
electrodynamics. Finally, our conclusions are presented in Section
4.

\paragraph{ K-essence and the Stephani Universe}

\noindent K-essence {[}10{]} is simply the most general model Lagrangian
density for a scalar field$\varphi$ involving its first covariant
derivative$\varphi_{,\mu}$ . We take the derivative to be time-like
so 
\begin{equation}
{\rm {\mathfrak{L}}}={\rm {\mathfrak{L}}}\left(\varphi,Y\equiv\sqrt{g^{\mu\nu}\varphi_{,\mu}\varphi_{,\nu}}\right)\label{2.1)}
\end{equation}
The use of the `velocity'\ $Y$ instead of the usual\ $X=g^{\mu\nu}\varphi_{,\mu}\varphi_{,\nu}$
as the kinematic variable considerably simplifies and clarifies the
subsequent treatment -- e.g. the stress-energy tensor takes the perfect
fluid form$\ T=\left(\rho+p\right)u_{\mu}u_{\nu}-g_{\mu\nu}p$ with
the identifications 
\begin{equation}
u_{\mu}={\varphi_{,\mu}\mathord{\left/{\vphantom{\varphi_{,\mu}Y}}\right.\kern -\nulldelimiterspace}Y}\,,\quad p={\rm {\mathfrak{L}}}\,,\quad\rho=Y{\rm {\mathfrak{L}}}_{,Y}-{\it {\mathfrak{L}}}\label{2.2)}
\end{equation}
Indeed, in co-moving coordinates\ $Y=\dot{\varphi}$ and\ $u_{\mu}={\delta_{\mu}^{0}\mathord{\left/{\vphantom{\delta_{\mu}^{0}N}}\right.\kern -\nulldelimiterspace}N}$
while the energy density is evidently just the Hamiltonian. The\ $\varphi$
field equation reads here 
\begin{equation}
\left({\rm {\mathfrak{L}}}_{,Y}g^{\mu\nu}{\varphi_{,\nu}\mathord{\left/{\vphantom{\varphi_{,\nu}Y}}\right.\kern -\nulldelimiterspace}Y}\right)_{:\mu}={\rm {\mathfrak{L}}}_{,\varphi}\label{2.3)}
\end{equation}
Imposing the nominal requirements of stability and causality the adiabatic
speed of sound squared is given by%
\footnote{ $For\ a\ derivation\, and\, further\, discussion\, see\,[11]$ %
} 
\begin{equation}
0\le c_{s}^{2}=\frac{p_{,Y}}{\rho_{,Y}}=\frac{{\rm {\mathfrak{L}}}_{,Y}}{Y{\rm {\mathfrak{L}}}_{,YY}}\le1\label{2.4)}
\end{equation}
That is to say the Lagrangian density must satisfy the inequalities$\ {\rm {\mathfrak{L}}}\ge0\;\&\;{\rm {\mathfrak{L}}}_{,YY}\ge0\;$
.

\noindent Particular classes of k-essence are factorizable models\ ${\rm {\mathfrak{L}}}\left(\varphi,Y\right)=-V\left(\varphi\right)F\left(Y\right)$
(which includes tachyon models {[}11, 12{]} for$F\left(Y\right)=\sqrt{1-Y^{2}}$
and the Chaplygin gas {[}13{]} in the subcase of constant potential)
and purely kinetic models (such as Scherrer's {[}14{]} model${\rm {\mathfrak{L}}}\left(Y\right)\simeq{\rm {\mathfrak{L}}}\left(Y_{0}\right)+{\rm {\mathfrak{L}}}''\left(Y_{0}\right){\left(Y-Y_{0}\right)^{2}\mathord{\left/{\vphantom{\left(Y-Y_{0}\right)^{2}2}}\right.\kern -\nulldelimiterspace}2}$
). Herein we are interested in models of the type {[}15{]} 
\begin{equation}
{\rm {\mathfrak{L}}}\left(\varphi,Y\right)=F\left(Y\right)-V\left(\varphi\right)=p\Rightarrow\rho=YF'\left(Y\right)-F\left(Y\right)+V\left(\varphi\right)\label{2.5)}
\end{equation}
This is because in co-moving co-ordinates$\varphi$ is a function
of\ $x^{0}=t$ only, so that the energy density will be homogeneous
if\ $YF'\left(Y\right)-F\left(Y\right)=0$ , i.e. for some constant$K$
\begin{equation}
{\rm {\mathfrak{L}}}\left(\varphi,Y\right)=KY-V\left(\varphi\right)\label{ZEqnNum703942}
\end{equation}
Note that in this linear velocity model the pressure is nonetheless
inhomogeneous via the lapse function\ $Y={\varphi_{,t}\mathord{\left/{\vphantom{\varphi_{,t}N\left(t,x^{i}\right)}}\right.\kern -\nulldelimiterspace}N\left(t,x^{i}\right)}$.
It is also notable that equations \eqref{1.2)} \eqref{2.3)} \& \eqref{ZEqnNum703942}
imply the Hubble expansion is directly related to the potential: $3KH=-{\partial V\mathord{\left/{\vphantom{\partial V\partial\varphi}}\right.\kern -\nulldelimiterspace}\partial\varphi}$
.

\paragraph{ K-essence and 5-D Non-linear ``Electrodynamics''}

\noindent Albeit the model \eqref{ZEqnNum703942} has the requisite
properties to serve as the source in the Stephani universe, one seems
to have traded one mystery for another: how is one to understand the
linear dependence on\ $Y$? To answer this we recall that long before
Kaluza and Klein Nordstrom {[}16{]} proposed to obtain a scalar gravity
theory from 5-dimensional ``electrodynamics'' by applying a ``cylinder
condition''. More specifically let the 5-dimensional co-ordinates
be denoted\ $x^{M}=\left(x^{\mu},y\right)$and for the 5-vector potential\ $A_{M}=\left(A_{\mu},\varphi\right)$;
assuming\ $A_{M}$is independent of the fifth co-ordinate the 5-dimensional
field strength\ $F_{MN}^{\left(5\right)}=A_{N,M}-A_{M,N}$ decomposes
as\ $F_{\mu\nu}^{\left(5\right)}=F_{\mu\nu},\; F_{\mu5}^{\left(5\right)}=\varphi_{,\mu}$.
Then taking the 5-dimensional metric\ $g_{MN}^{\left(5\right)}$
of the form\ $g_{\mu\nu}^{\left(5\right)}=g_{\mu\nu},\; g_{\mu5}^{\left(5\right)}=0,\; g_{55}^{\left(5\right)}=-1$we
have 
\begin{equation}
-\frac{1}{2}F^{\left(5\right)}\cdot F^{\left(5\right)}\equiv-\frac{1}{2}F_{MN}^{\left(5\right)}F^{\left(5\right)MN}=-\frac{1}{2}F_{\mu\nu}F^{\mu\nu}+g^{\mu\nu}\varphi_{,\mu}\varphi_{,\nu}=-\frac{1}{2}F\cdot F+Y^{2}\label{2.7)}
\end{equation}
As\ $\left[A\bullet A\right]^{\left(5\right)}=A_{M}A^{M}=A_{\mu}A^{\mu}-\varphi^{2}=A\bullet A-\varphi^{2}$,
for compact\ $y$\ it follows that \textit{any} k-essence model$\ {\mathfrak{L}}\left(\varphi,Y\right)$can
be obtained from a 5-dimensional model%
\footnote{ $The\ appearance\ of$ $\left[A\bullet A\right]^{\left(5\right)}$
$as\ such\ breaks\ gauge\ invariance,\ but\ can\ be\ understood\ as\ the\ unitary\ gauge\ limit\ of$
$\left[A^{\left(\theta\right)}\bullet A^{\left(\theta\right)}\right]^{\left(5\right)}$
$with$ $A_{M}^{\left(\theta\right)}=A_{M}+\theta_{,M}$ $.$ %
}${\rm {\mathfrak{L}}}^{\left(5\right)}\left(\sqrt{-\left[A\bullet A\right]^{\left(5\right)}},\sqrt{-\frac{1}{2}F^{\left(5\right)}\cdot F^{\left(5\right)}}\right)$by
dimensional reduction provided also we set\ $A_{\mu}=0$.

\noindent Taking the range of the fifth co-ordinate as\ $0\le y\le l_{5}$,
for our model source in the Stephani universe 
\begin{equation}
{\rm {\mathfrak{l}}}_{5}{\rm {\mathfrak{L}}}^{\left(5\right)}=K\sqrt{-\frac{1}{2}F^{\left(5\right)}\cdot F^{\left(5\right)}}-V\left(\sqrt{-\left[A\bullet A\right]^{\left(5\right)}}\right)\label{ZEqnNum493458}
\end{equation}
Similar kinetic terms appear in the context of Born-Infeld electrodynamics
and D-branes in string/M-theory. Of particular note is that Nielson
and Oleson {[}17{]} proposed a Lagrangian density of the form$\sqrt{-\frac{1}{2}F\cdot F}$as
a field theory for dual strings. We thus suggest that the kinetic
part of \eqref{ZEqnNum493458} be understood as originating in string/M-theory.

\paragraph{ Conclusions}

\noindent In this paper we have considered the issue of the matter
source in the Stephani universe as an inhomogeneous alternative to
the FRW model with a cosmological constant. We have shown that a form
of k-essence has the requisite properties to be that source, and that
this k-essence can be obtained by dimensional reduction of a 5-dimensional
model reminiscent of an early field theory for strings.

\paragraph{Acknowledgements}

\noindent This work was supported by a grant from the National Research
Foundation.

\paragraph{References}
\begin{enumerate}
\item \textbf{ }S. Perlmutter et al., Astrophys. J. \textbf{517}, 565 (1999);
A. G. Riess et al., Astron. J. \textbf{116}, 1009 (1998). 
\item S Weinberg, Rev. Mod. Phys. 61, 1 (1989) 
\item C Clarkson, G Ellis, J Larena and O Umeh, Rept. Prog. Phys. 74, 112901
(2011) {[}arXiv:1109,2314{]} 
\item M Redlich, K Bolejko, S Meyer, G F Lewis and M Bartelmann, Astron.
Astrophys. \textbf{570}, A63 (2014) {[}arXiv:1408.1872{]} 
\item J Adamek, C Clarkson, R Durrer and Martin Kunz, arXiv:1408.2741 
\item R C Tolman, \textit{Relativity, Thermodynamics and Cosmology}, Oxford
University Press, Cambridge, England 1934 
\item M Wyman, Phys. Rev. \textbf{70}, 396 (1946) 
\item H Stephani, Commun. Math. Phys. 4,~137 (1967) 
\item S Sedigheh Hashemi, S Jalalzadeh and N. Riazi, arXiv:1401.2429 and
references therein 
\item C Armendariz-Picon, V Mukhanov, and P J Steinhardt, Phys. Rev. Lett.
\textbf{85}, 4438 (2000); astro-ph/0004134 
\item N Bilic, G B Tupper and R D Viollier R D, Phys. Rev. \textbf{D80},
023515 (2009) {[}arXiv:0809.0375v3{]} 
\item A Sen Mod. Phys. Lett. \textbf{A 17}, 1797 (2002); J. High Energy
Phys. \textbf{04}, 048 (2002); \textbf{07}, 065 (2002). 
\item N Bilic, G B Tupper and R D Viollier, Phys. Lett. \textbf{B535}, 17
(2002) 
\item R J Scherrer Phys. Rev. Lett. \textbf{93}, 011301 (2004) 
\item J De-Santiago, J L Cervantes-Cota and D Wands, Phys. Rev. \textbf{D87},
023502 (2013) 
\item G Nordstrom, Phys. Zeitschr. \textbf{15}, 504 (1914) 
\item H B Nielsen and P Oleson, Nucl. Phys. \textbf{B57}, 367 (1973) \end{enumerate}

\end{document}